\documentclass{aa}
\usepackage{amssymb}
\usepackage{times}
\usepackage{graphicx}
\usepackage[english]{babel}
\usepackage{hyperref}

\def\asec{\ifmmode ^{\prime\prime}\else$^{\prime\prime}$\fi}

\def\msun{M$_{\odot}$}

\def\it{\sl}
\def\degs{\ifmmode ^{\circ}\else$^{\circ}$\fi}
\def\amin{\ifmmode ^{\prime}\else$^{\prime}$\fi}
\def\asec{\ifmmode ^{\prime\prime}\else$^{\prime\prime}$\fi}
\def\fd{\hbox{$.\!\!^{\rm d}$}}            % Fractions of days
\def\wga{1\,WGA\,J1958.2+3232\,}

\begin{document}
\thesaurus{06.                  % A&A Section 6: Formation, structure and 
evolution of stars
                (08.14.2;       % Stars: novae, cataclysmic variables
                08.09.2;        % Stars: individual
                03.20.4;        % Techniques: photometric
                03.20.9)        % Techniques: spectroscopic
                }
\title{ The   Orbital Period of  Intermediate Polar 1WGA J1958.2+3232} 

\author{Sergei. V. Zharikov
\and  Gaghik. H. Tovmassian
\and Juan Echevarr\'{\i}a 
\and Aixa Aub\'{e} C\'{a}rdenas}

\institute{ Observatorio Astron\'{o}mico Nacional, Instituto de Astronom\'{i}a, UNAM,\\ 
22800, Ensenada, B.C., Mexico\thanks{use for smail P.O. Box 439027, San Diego, CA, 92143-9027, USA}
}
\authorrunning{S. V. Zharikov et al., }
\titlerunning{The Orbital Period of  1WGA J1958.2+3232} 
\offprints{Zharikov S}
\mail{zhar@astrosen.unam.mx}
\date{Received --  / Accepted -- }
\maketitle

\begin{abstract}
The detection of the  orbital period of $4\fh36$ is reported for the new
Intermediate Polar \object{\wga}.  The orbital period was derived from
time-resolved   photometric  and   spectral  observations.    We  also
confirmed the $733$ sec spin period of the White Dwarf consistent with
the X-ray pulsations  and were able to distinguish  the beat period in
the light  curve.  Strong modulations with orbital  period are detected
in the  emission spectral lines from spectral  observations. They show
the presence  of a bright hot  spot on the  edge of the  accretion disk.
The  parameters of  this  recently discovered  Intermediate Polar  are
determined.


\end{abstract}

\section{Introduction.}

   Cataclysmic variables (CVs) are  close binary systems in which mass
is transferred from a red dwarf  star that fills its Roche lobe onto a
white  dwarf (WD).  Intermediate  polars  (or DQ  Her  systems) are  a
subclass  of  magnetic  cataclysmic  variables with  an  asynchronously
rotating     ($P_{spin}<    P_{orb}$)    magnetic     white    dwarf
(Patterson, \cite{Patterson2}; Warner, \cite{Warner}).  The accretion flow from the red dwarf
star forms an  accretion disk around white  dwarf, and this  disk is
disrupted by the magnetic field  close to the white dwarf.  Within the
magnetospheric radius, the material  is channeled towards the magnetic
polar regions in an arc-shaped accretion  ( Rosen,\cite{Rosen})

The   recently  discovered   pulsating   X-ray  source   \object{\wga}
(Israel et al., \cite{Israel1}) was  announced as a  new Intermediated Polar  (IP) by
Negueruela et al., (\cite{Negueruela}) from  spectral observations.  Strong  modulations of
this source  in X-rays were obtained  from ROSAT PSPC ($721\pm14$  sec) and
more accurate $734\pm1$ sec from ASCA are presented by Israel et al. (\cite{Israel1}) and 
Israel et al. (\cite{Israel2}). Photometric observations of the optical counterpart of
\object{\wga}  exhibited strong optical variations,  compatible with the  X-rays ($\sim
12$ min) period ( Uslenghi  et al., \cite{Uslenghi}).  This modulation was interpreted as
evidence of a spin period of the WD in this close binary system.

In  this paper  we  present the  results  of   new photometric  and
spectral observations of this system.

\section{Observations.}

The  CCD photometric  and spectral  observations of  the \object{\wga}
were carried out on 2--5 August  2000 at the 1.5m and 2.12m telescopes
of the  Observatorio Astronomico Nacional, San Pedro Martir  of the
Institute of  Astronomy of  UNAM, Mexico. The  log of  observations is
presented in Table.\ref{logObs}.

\begin{table}[t]
\caption{Observations Log}
\begin{tabular}{lcccc}
\hline\hline
HJD start &   Duration  & Time  & Band          & Telescope   \\
day     &           & of exposure      &                &            \\  
254100+  & min       &  sec             &               &             \\ \hline
759.719$^a$    & 380       &   120            &  $R_c$        &  1.5m       \\
760.671         &  435       &   120            &  $R_c$        &  1.5m       \\
761.855         &  173       &   700            &  4025-5600\AA &  2.12m      \\
762.649         &  461       &   700            &  4025-5600\AA &  2.12m      \\
763.678         &  319       &   700/350        &  4025-5600\AA &  2.12m      \\ \hline 
\end{tabular}
\label{logObs}
\begin{tabular}{ll}
$^a$ 2 August. \\
\end{tabular}
\vspace{-0.5cm}
\end{table}
%\begin{table*}[t]
%\caption{Observations Log}
%\begin{tabular}{lccccc}
%\hline\hline
%HJD             &   UT start & Duration  & Time of exposure & Band          & Telescope   \\
%day             &    h       & min       &  sec             &               &             \\
%2541759 (2 Aug) &   05:14    & 380       &   120            &  $R_c$        &  1.5m       \\
%2541760         &   04:06    & 435       &   120            &  $R_c$        &  1.5m       \\
%2541761         &   08:31    & 173       &   700            &  4025-5600\AA &  2.12m      \\
%2541762         &   03:35    & 461       &   700            &  4025-5600\AA &  2.12m      \\
%2541763         &   04:16    & 319       &   700/350        &  4025-5600\AA &  2.12m      \\ \hline 
%\end{tabular}
%\label{logObs}
%\end{table*}

\begin{figure}[t]
\setlength{\unitlength}{1mm} 
\resizebox{12cm}{!}{ 
\begin{picture}(180,110)(0,0) 
\put (0, 0){\includegraphics[width=135mm]{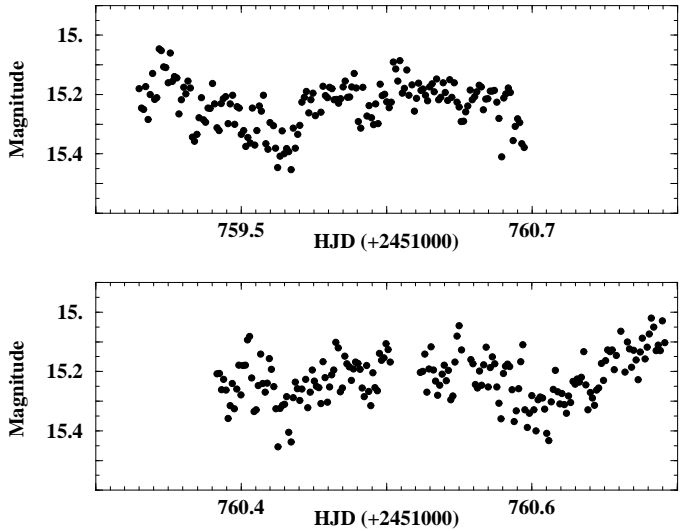}}
%\put (140, 0){\includegraphics[width=135mm]{Power.eps}}   
\end{picture}
}\hfill 
\caption{ \object{\wga} light  curves in Rc band are presented.
Binning time is 120s.}
\label{light}
\end{figure}
\begin{figure}[t]
\setlength{\unitlength}{1mm} 
\resizebox{12cm}{!}{ 
\begin{picture}(180,120)(0,0) 
%\put (0, 0){\includegraphics[width=135mm]{phot.eps}}
\put (0, 0){\includegraphics[width=135mm]{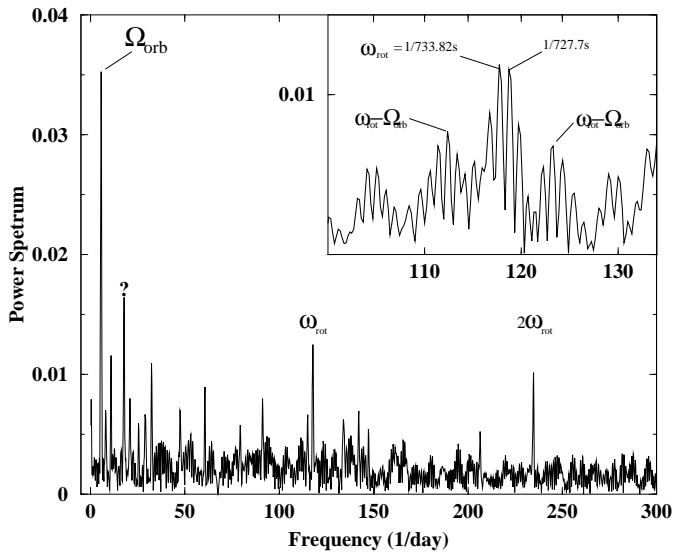}}   
\end{picture}
}\hfill 
\caption{The  CLEANed power spectrum of Rc light curve
is given.  The orbital  period $\Omega_{orb}$ and $\omega_{rot}$ spin
period frequencies are marked.  In the inset  part of the  cleaned power
spectrum  with  a fewes   number   of  iterations  is  shown.   The  beat
frequencies  $\omega_{rot}-\Omega_{orb}$,  $\omega_{rot}+\Omega_{orb}$
and $\omega=1/727.78$ sec  are presented.}
\label{clean}
\vspace{-0.5cm}
\end{figure}

\subsection{Optical photometry}

We  obtained  $R_c$-band   time-resolved  photometry  of  the  optical
counterpart of \object{\wga}  during two nights in August  2000 at the
1.5m telescope.  The telescope  was equipped with a $1024\times 1024$\
pixel SITE CCD. The frame was  reduced in size to { $450 \times 450$\,
pix} for faster read-out.  It accommodated the object and at least two
comparison stars  in the field of  view.  The exposure  times were 120
sec,  which leads  to  the time  resolution  of 169  sec, taking  into
account dead time between readouts.  In total the object was monitored
during {$\sim$13.55h (6.3h the  first night and 7.25h the second).}
The data reduction was performed by using ESO-MIDAS\footnote{ESO-MIDAS
is  a copyright protected  software product  of the  European Southern
Observatory, and provides general  tools for image processing and data
reduction.}  and IRAF
\footnote{IRAF is the Image Reduction and Analysis Facility, a general
purpose software system for the reduction and analysis of astronomical
data.  IRAF is written and  supported by the IRAF programming group at
the  National  Optical   Astronomy  Observatories  (NOAO)  in  Tucson,
Arizona.   NOAO is  operated by  the Association  of  Universities for
Research in  Astrono my (AURA), Inc. under  cooperative agreement with
the National Science Foundation} softwares.  The images were corrected
for bias and flatfield before aperture photometry was carried out.  An
estimate  of  uncertainty  of   the  CCD  photometry  of  the  optical
counterpart  of  \object{\wga} was  obtained  from  the dispersion  of
magnitudes in the differential photometry of comparison stars with 
similar brightness.  The dispersion ranged from 0.005 to 0.01 mag.  We
did not obtain an absolute calibration for our photometric data.

\subsection{Optical spectroscopy}

Time-resolved spectroscopy of the  optical counterpart of
\object{\wga}  was obtained on   4-6  Aug,  2000   using  the  Boller   \&  Chivens
spectrograph installed in the Cassegrain focus of the 2.12m telescope.
We used  the 400\,l/mm grating  with a $13^o54$ blaze  in the  second order,
combined with  the blue BG39 filter and  CCD TEK$1024\times 1024$\,pix
with a $0.24\mu$ pixel size.  The slit width was 1.5 arcsec projected on
the sky.  This combination yielded a spectral resolution of $2.7$\AA \
FWHM and provided a wavelength  coverage of $\lambda 4050- 5600$ \AA \.
  From three nights of spectral observations the second and third
nights were  disrupted by passing  clouds.  However the  seeing was
satisfactory  with  images being  $\leq  1.2$  arcsec.   The slit  was
oriented with  position angle of  {$306^o$} to accommodate   nearby
star for the  flux level control.  The exposure time  in the first two
nights was 700 sec, while on the  third night,  700 and 350 sec.  The
He-Ar comparison spectra  were taken every $\sim 120\  $ min.  A total
of  68  spectra  were  obtained.   The IRAF  long  slit  spectroscopic
reduction package  was used for extraction of  spectra, wavelength and
flux calibrations.  Before  head the images were reduced  for bias and
cosmic rays.

\section{Results}

\subsection{Light Curve. Orbital and Spin Modulations}

The  object demonstrates  multi-scale  time variability  in the range  0.3
magnitudes (see. Fig.\ref{light}).  Four pronounced eclipse--like depressions obviously shape
the  light  curve.  Strong  flickering with  optical  pulse  amplitude
(semiamplitude) of  about 0.1 magnitude is also obvious in the light curve
detected and  identified earlier (Israel et al., \cite{Israel1}, Uslenghi et al., \cite{Uslenghi}) as
spin related modulations.  The photometric data of \wga \ was analyzed
for  periodicities   using  Discrete  Fourier   Transform  (DFT)  code
(Deeming, \cite{Deeming}) with the  CLEAN procedure (Roberts et al., \cite{Roberts}).   The CLEANed
power spectrum (Fig.\ref{clean}) of Rc photometric data shows a
clear peak at {\bf $\Omega_{orb}=5.54996\pm0.39624$}, corresponding to
$P=0.1802\pm0.0065d$. This peak is  caused by above mentioned eclipses
in the light curve and clearly marks the orbital period of the system.

We found also a significant  peak at the spin period $\omega_{rot}$ of
the  WD  corresponding to  $733.82\pm  1.25$sec.   This  period is  in
excellent  accordance  with  that  recently discovered  by  ASCA  X-ray
pulsations     (Israel et al., \cite{Israel2}).     The    beat     frequencies    at
$\omega_{rot}-\Omega_{orb}$,   $\omega_{rot}+\Omega_{orb}$   are  also
present  in  the  CLEANed  power  spectrum but  with  a fewer  number  of
iterations   (see  insert   in   the  upper   right   corner  of   the
Fig\,\ref{clean}).  The harmonics  of basic frequencies $\Omega_{orb}$
and  $\omega_{rot}$ are detected  as well.   Besides these,  there are
comparably significant  peaks at the periods of  727.78\,sec and 1.36h.  The
former was  detected also by  Uslengi et al., (\cite{Uslenghi}) and is probably  the one
day alias  of the  $\omega_{rot}$, while for  the latter we  could not
find any reasonable explanation.

\begin{figure}
\begin{center}
\includegraphics[width=85mm]{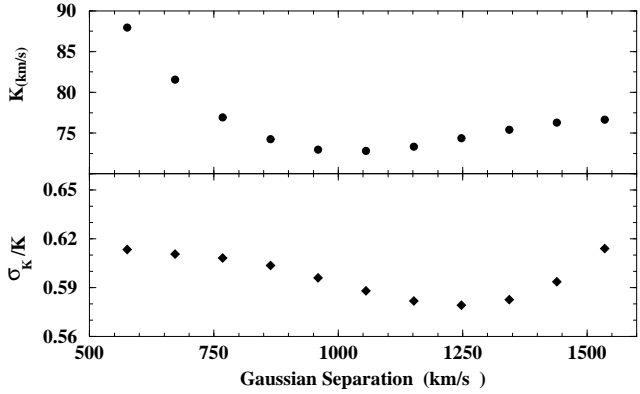}   
\end{center} 
\caption{The   diagnostic  diagrams   for  the   $H_{\beta}$  emission
lines. The radial velocity  semi-amplitude K, the ratio $\sigma_{K}/K$
are plotted as  a function of the Gaussian  separation, obtained for a
period of 0.18152 days.}
\label{diagnos}
\end{figure}

\begin{figure}
\begin{center}
\includegraphics[width=85mm]{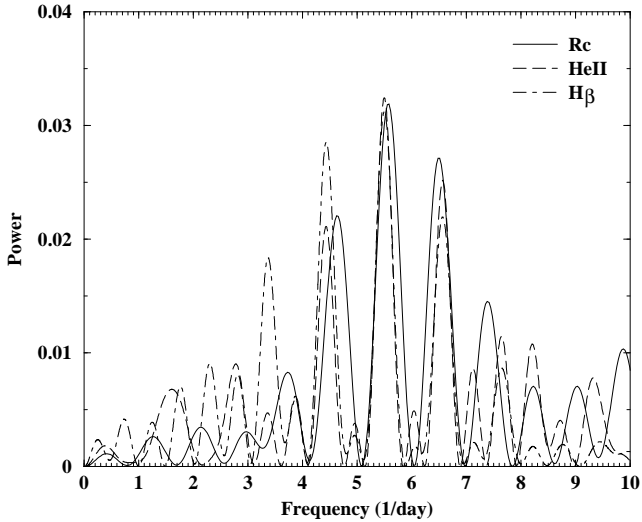}   
\end{center}
\caption{The  power  spectrum   of the  $R_c$ light curve and the  \ion{He}{ii} 4686 and  \ion{H$_{\beta}$}{}
radial velocity curves  are presented.  They are scaled to
the amplitude of Rc power spectrum. The maximum peak of frequency corresponds
to  the orbital period of the system  $P_{orb}=0.18152\pm0.0011d$ .}
\label{powSpec}
\vspace{-0.5cm}
\end{figure}

\begin{figure}
\includegraphics[width=85mm]{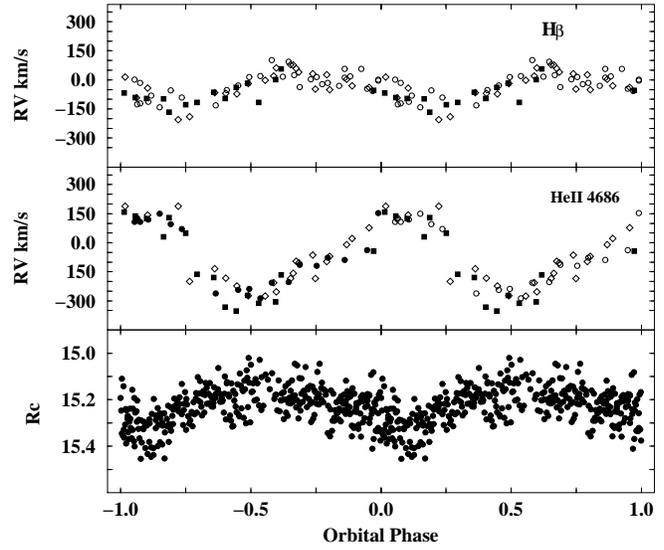}   
\caption{The  radial velocity curves for the emission
lines of \ion{H$_\beta$}{} and  \ion{He}{ii} 4686 phased  on the  spectroscopic orbital
period ($0.18152$   days) are presented  upper and middle panels respectively.
 The filled  circles corresponds to 4  Aug,
filled rectangulars  and open deamons  are 5 and  6 Aug,  respectively.  Binning
time is 700s.  Rc light curve  of \object{\wga}  is presented in the lower panel.
The  binning time is 120sec.}
\label{period}
\vspace{-0.5cm}
\end{figure}

\subsection{Radial Velocity Variations and Binary System Parameters}

The spectrum  of \object{\wga}  shows features characteristic  to Cataclysmic
Variables.  We  refer to  the Fig.1  of  Negueruela et al., (\cite{Negueruela}), who
obtained  spectrum of  the object  in a  wider spectral  range and  with
better spectral resolution.   Meanwhile we obtained  time-resolved spectroscopy
of  \object{\wga}    around  the emission  lines of  \ion{H$_{\beta}$}  and 
\ion{He}{ii} ,  covering several  orbits.  Thus, we were able  to
consider  periodical  variations  in   the  spectrum  of  the  object,
primarily in  the emission lines.  The simple  stacking of consecutive spectra
 onto  the  trailed  spectrum  showed  strong  variability  in lines. It is
distinct in Balmer lines and in the  higher excitation lines of ionized  Helium.  
The Balmer  lines  are  double--peaked with  S-wave moving inside,  which
makes it hard  to see the periodic  pattern.  In the \ion{He}{ii} 4686 line the
central narrow component dominates in most of the phases, and it shows clear
sinusoidal variation.

In  order to  determine the  orbital elements  we measured  the radial
 velocities (RV)  of  \ion{H$_{\beta}$}{}    applying  the  double  Gaussian
 deconvolution method introduced by Schneiger \& Young, (\cite{Schneider}) further developed
 by  Shafter (\cite{Shafter}).   This  method   is  especially   efficient  for
 measurements of the orbital motion  of CVs with a prominent spot at the
 edge of  the accretion disk,  contaminating the central parts  of the
 emission lines. It allows us to measure RV variations using the wings of the
 lines.  The width of Gaussians were chosen to be slightly larger than
 our spectral resolution (8.5 \AA), where deconvolution was reached at
 all  orbital  phases.   The  radial  velocities were  measured  as  a
 function of  distance {$ a  $ } between  the Gaussians, and  then the
 diagnostic diagrams  were constructed using an initial  guess for the
 orbital  period, derived  from photometry  and from  preliminary radial
 velocity measurements  via Gaussian fits  to the lines.   The optimal
 value  of separation ($a=  1175$ km  /sec )  was determined  from the
 diagnostic diagrams,  and the RV  values measured for  these Gaussian
 separations  were again  subjected to  a power  spectrum  analysis in
 order  to  refine the  period.  The  spectroscopic  period peaked  at
 slightly longer value, than the photometric (however within errors of the
 photometric  period). This  method  quickly converged  and after  two
 iterations  no  further  improvement  was  achieved.  The  diagnostic
 diagrams for  \ion{H$_{\beta}$}{}   are shown in  Fig.\ref{diagnos}.
Fig.\ref{period} (top) shows  the \ion{H$_{\beta}$}{} radial velocity curve.

\begin{figure}[t]
\includegraphics[width=85mm]{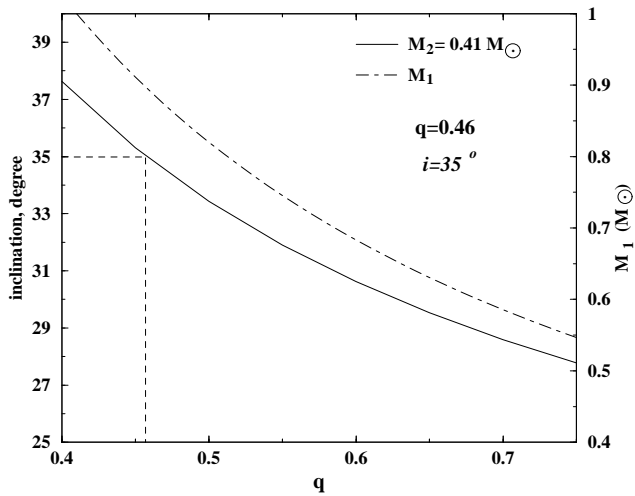}   
\caption{The angle of inclination of the system vs. $q=M_2/M_1$ 
are shown (solid line).The mass of WD in the system  M$_1$ vs. $q$ 
 for $M_2=0.41M_{sun}$ are given by the 
dot-dashed line. ``Best fit''  parameter from the secondary star position and the ballistic trajectory of the gas stream on the Dopplers maps are marked. 
}
\label{angle}
\end{figure}
A narrow single Gaussian profile  was fitted to the prominent emission
features  in the  profile  of  \ion{He}{ii}  $\lambda  4686$\AA, and  the
measured  line centers  were  used to  determine  the radial  velocity
solution  for  \object{\wga}.  The radial  velocity  curve  for  \ion{He}{ii}
$\lambda4686$  line  is   presented   on   the   middle  panel   of
Fig.\ref{period}. These  measurements were also subjected to  the power spectrum
analysis.   The obtained  orbital  period is  in  good agreement  with
the values   derived   from the   photometry   and  the  \ion{H$_\beta$}{} line   radial
velocities. The  power spectra around the values  corresponding to the
orbital period from these three independent determinations are plotted
in the Fig.\ref{powSpec}. One  can see the excellent match of the central peak.
We adopted the  $0\fd 18152\pm 0\fd 00011$ as the  final value for the
orbital  period  of  \object{\wga}  from  our observations.  Longer  time  base
observations are needed to improve this value.

    Each   of  the  radial   velocity  curves   was  fitted   using  a
least-squares routine of the form
\begin{equation}
v(t) = \gamma_{o}+K_{1}*\sin(2\pi(t-t_o)/P),
\end{equation}
where $\gamma_o$ is  the systematic velocity of the  system, and K$_1$
is the  semi-amplitude of radial  velocity, both in km  $s^{-1}$.  The
observation time is $t$,  the epoch $t_0=2451761\fd2281\pm0\fd0001$
corresponds to the +/- zero  crossing of the $H_\beta$ radial velocity
curve,  and therefore  is a  superior  conjunction  of  the binary  system
(secondary  located between  observer  and the  WD).  Accordingly  the
phase 0.0  was calculated  and operated further   at this
epoch.  Table.\ref{RV}  gives summary of the radial  velocity fits for
\ion{H$_{\beta}$}{} and \ion{He}{ii} 4686 emission line.

\begin{table}[h]
\caption{Radial velocity solution parameters}
\begin{tabular}{ccc}
\hline\hline
Name         & $\gamma_o$             &   $K_1$          \\  
             &  $km/s$                &   $km/s$               \\ \hline
$H_{\beta}$  & $-39\pm5$              & $74\pm7$                      \\
$He II 4686$ & $-70.2\pm 15.0$           & $ 197.4\pm 25.7$        \\ \hline
\end{tabular}
\label{RV}
\end{table}

After refining the orbital period from spectroscopy, and determining the phase 0.0,
the photometric light curve was folded by the corresponding parameters and presented
in the lower panel of Fig.\ref{period}. 

\begin{figure*}[t]
\setlength{\unitlength}{1mm} 
\begin{picture}(175,155)(0,3) 
\put ( 0,11){\includegraphics[width=80mm,bb=-80 -20 385 414,clip]{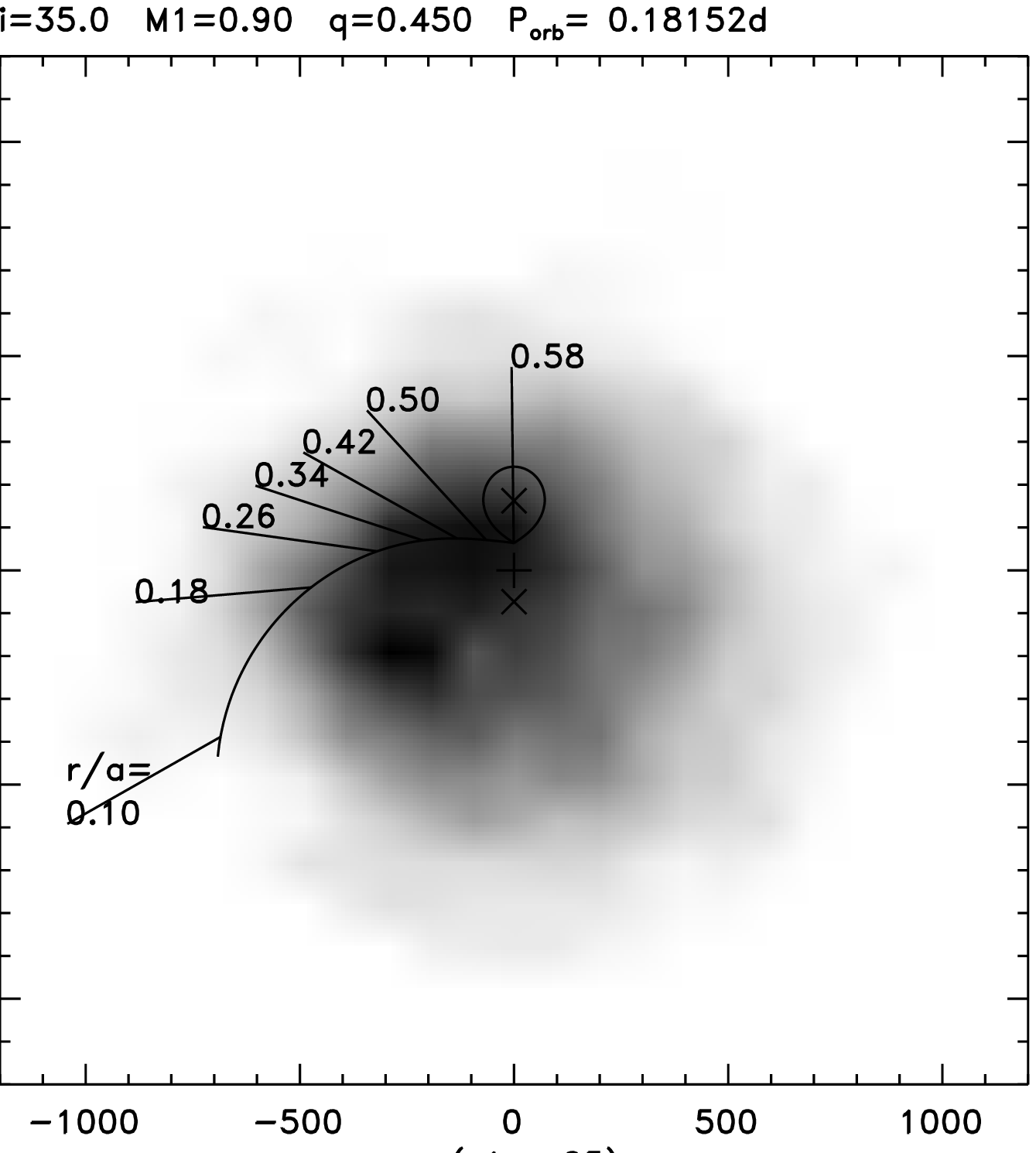}}
\put ( 0,88){\includegraphics[width=72mm,bb=70 500 360 700,angle=90,clip=]{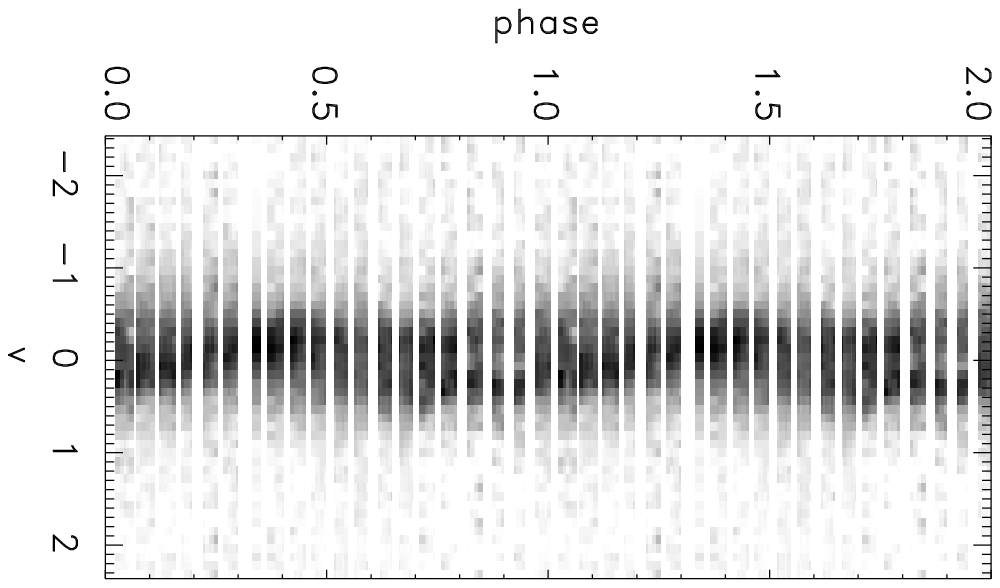}}   
\put (45,88){\includegraphics[width=72mm,bb=70 500 360 655,angle=90,clip=]{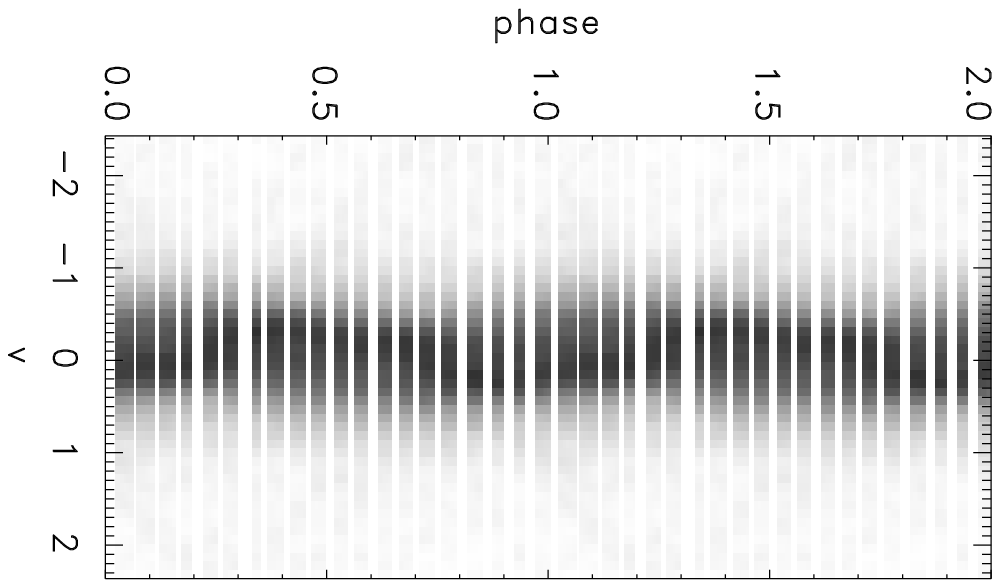}}
 
\put (82,11) {\includegraphics[width=80mm,bb=-80 -20 385 414,clip]{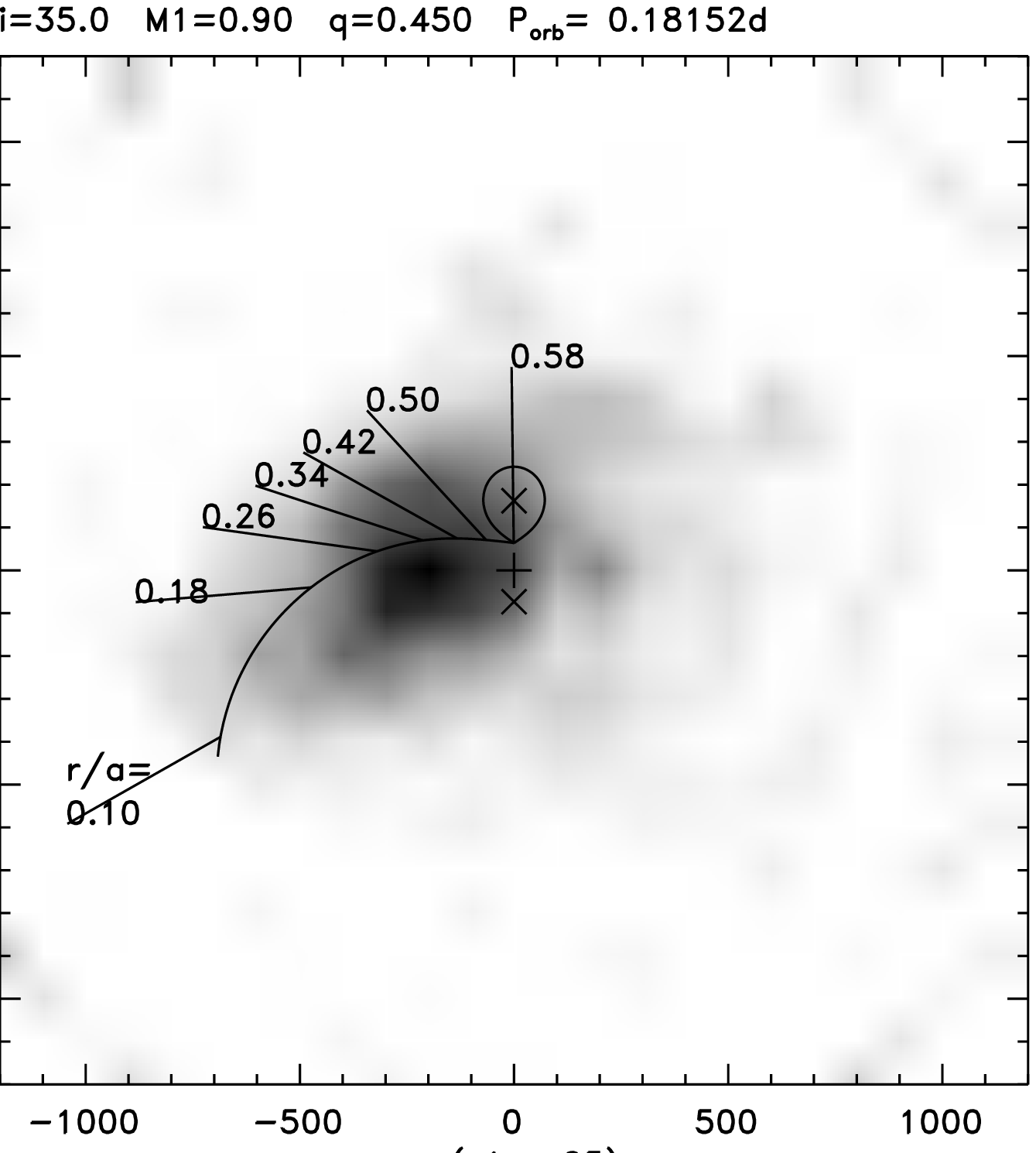}}
\put (84,88){\includegraphics[width=72mm,bb=70 500 360 700,angle=90,clip=]{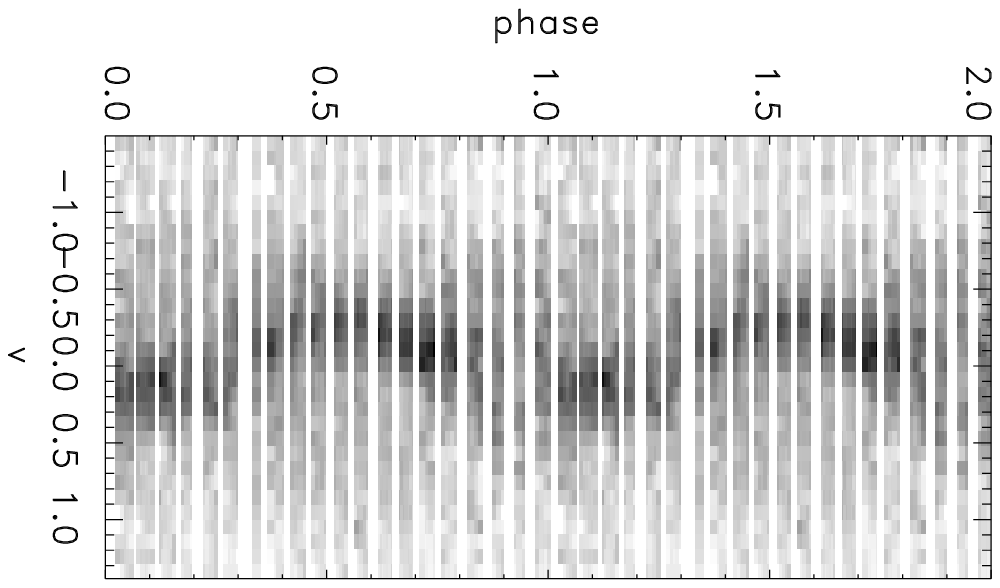}}   
\put (128,88){\includegraphics[width=72mm,bb=70 500 360 655,angle=90,clip=]{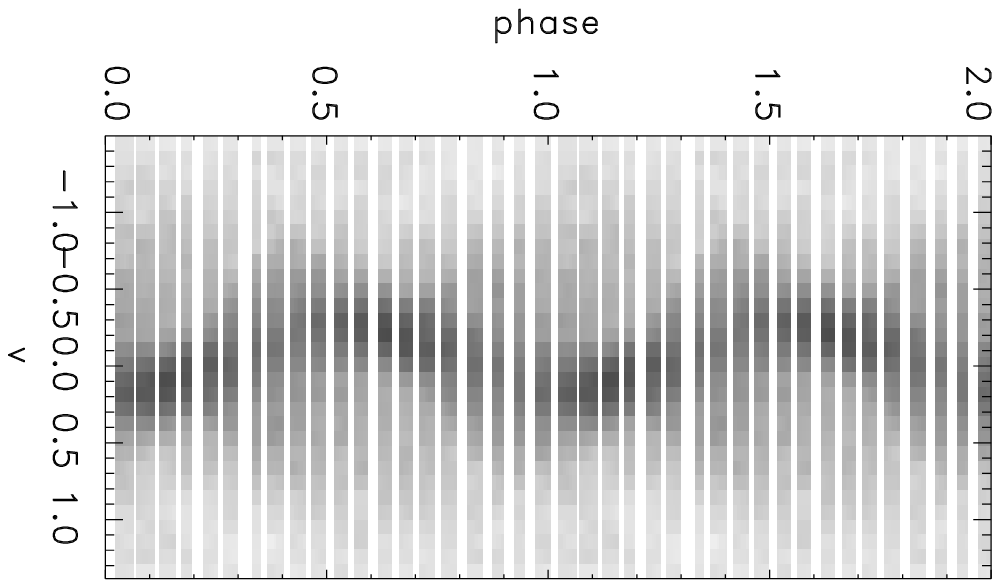}}
\end{picture}
\caption{ Trailed,   continuum-subtracted,   spectrums    of   1WGA
J1958.2+3232 plotted in two cycles. Doppler maps of the emission lines
\ion{He}{II}  (left   panel),  \ion{H$_\beta$}{}  (right  panel)   in 
velocity  space ($V_x,V_y$) are  given.  A schematic  overlay marks the Roche 
lobe of the secondary, the ballistic  trajectory and the magnetically funneled
horizontal  part of  the  accretion stream.   The  secondary star  and
gas-stream trajectory are plotted for K=74 km/s and q=0.46. }
\label{doppler}
\end{figure*}

Several  conclusions can  be made  after considering the  three  curves in
Fig.\ref{period}   in conjunction with  and taking  into account  common  knowledge of
Intermediate Polar systems (Patterson, \cite{Patterson2}, Warner, \cite{Warner}):
\begin{itemize}
\item The double  peaked Balmer  lines  are due  to  the  presence of  an
accretion disk or  ring orbiting the primary  WD in  this IP. RV variations
in  the wings of  lines describe the  rotation of  the primary  of  the binary
system.
\item The S-wave present in the Balmer  and \ion{He}{ii} emission lines 
is  evidence for a  hot compact region  on the accretion disk.
\item  From  the  difference   of  phases  and  amplitudes  of  radial
velocities of  wings of \ion{H$\beta$}{}  and the narrow  component of
\ion{He}{ii} we can localize the hot spot in the second quarter of the
disk,  if the center  of coordinates  is at  the WD.  It is  the usual
location  of the  spot originating  from the  impact of the  mass transfer
stream with the accretion disk. However it is also the area which most
commonly  heated by  the energetic  X-ray beam  from  the magnetically
accreting pole  on the  surface of the  WD in  IPs (see the  sketch in
Hellier et al., \cite{Hellier}).
\item The hot  spot is self eclipsed by the  accretion ring as follows
from the  phasing of  the light  curve, the dips  in the  light curve
centered at  the phase $\approx  0.12\phi_{orb}$. The primary  at this
phase starts  to move toward the observer  and the hot spot  is on the
opposite side of the ring approaching maximum velocity.
\end{itemize}

We  estimate  the mass and radius of the secondary star as $M_2 = 0.41$\msun
and $R_2=0.47R_{\sun}$ from
\begin{eqnarray}
M_2/M_{sun} & = & 0.0751 P(h)^{1.16}, \nonumber \\  
R_2/R_{sun} & = &  0.101P(h)^{1.05}, \   1.4< P(h)<12 
\end{eqnarray}
 the mass-period and radius-period
relations of Echevarr\'{i}a, (\cite{Echevarria})

The mean mass estimate of 76 CV white dwarfs is $M_{WD} = 0.86M_{\sun}$ (Sion, \cite{sion}). 
Webbink, (\cite{webbink}) gives statistically  average white dwarf masses 
ratios (q = 0.29) and average masses for all systems ($M_{wd} = 0.61 M_{\sun}$)
 below the period gap and ($q = 0.64$,$M_{wd} = 0.82 M_{\sun}$)  above the period gap.

 From  our spectroscopic  radial velocity  solution, we  can determine
preliminary values  for the basic system  parameters of \object{\wga}.
The  binary orbital  plane  inclination can  be  determined from  (see
Downes et al., \cite{Downes}; Dobrzycka \& Howel, \cite{Dobrzycka}):
\begin{equation}
  \sin^3(i)=\frac{K_1^3 P}{2\pi G M_2}(\frac{q+1}{q})^2
\end{equation}
if the mass ratio  of the system is known. 
The dependance of $i$ versus $q  = M_2/M_1$ in the range
0.4 up to 0.75 is shown in Fig.\ref{angle} for the above determinated  ${K_1}_{H_{\beta}}$.

We attempted  to refine this estimate for  \object{\wga} by constraining
Doppler tomograms from observed emission line profiles.
\begin{table}[t]
\caption{The 1WGA1958.3+3232 adopted system parameters}
\label{Param}
\begin{tabular}{cccc}
\hline\hline
Parameter  & Value        &  Parameter & Value            \\  \hline
$P_{orb}$  & 0.18152d     & $R_2$      & 0.47 $R_{\sun}$  \\
$P_{rot}$  & 733.7s       & $a$        &  1.5 $R_{\sun}$  \\
$M_2$      &0.41$M_{\sun}$& $i$        & $35^o$           \\
$q$        & 0.46         & $M_1$      & 0.9$M_{\sun}$     \\ \hline
\end{tabular}
\end{table}
\begin{figure}[t]
\setlength{\unitlength}{1mm} 
\begin{picture}(175,80)(0,3) 
\put (0,11) {\includegraphics[width=80mm,bb=-80 -20 385 414,clip]{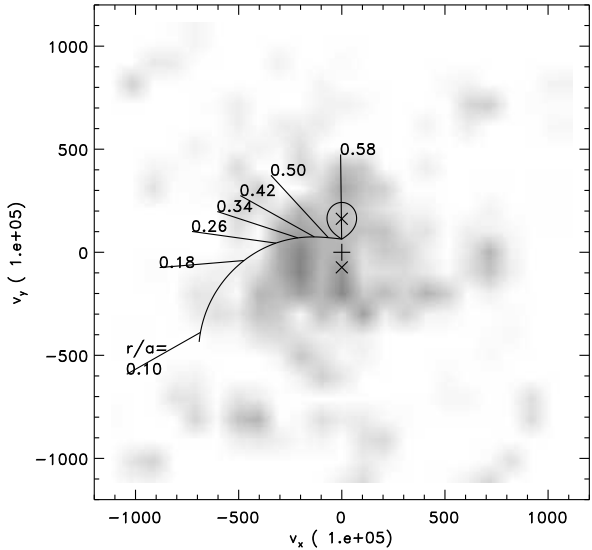}}
\end{picture}
\caption{Doppler maps of  the emission line blend \ion{C}{iii}/\ion{N}{iii}
 in  velocity space ($V_x,V_y$)  is given.  A  schematic overlay
marks the  Roche lobe of  the secondary, the ballistic  trajectory and
the magnetically funneled horizontal part of the accretion stream. The
secondary star and gas-stream trajectory are plotted for K=74 km/s and
q=0.46. }
\label{doppler1}
\end{figure}
Doppler tomography is a useful  tool to extract further information on
CVs  from  trailed  spectra.   This  method,  which  was  developed  by
Marsh \& Horne, (\cite{Marsh}),  uses the  velocity  profiles of  emission-lines at  each
phase to  create a  two-dimensional intensity image  in velocity-space
coordinates ($V_x,V_y$).
 Therefore,  the  Doppler  tomogram  can  be  interpreted  as  a
projection of emitting regions in cataclysmic variables onto the plane
perpendicular to  the observers view.   We used the code  developed by
Spruit (\cite{spruit}) to  constrain Doppler  maps of  \object{\wga} with
maximum entropy  method.  The resulting Doppler maps  (or tomogram) of
emission  lines   of  \ion{H$\beta$}{},  \ion{He}{ii}   and  the blend  of
\ion{C}{iii}/\ion{N}{iii}  are  displayed  as  a gray-scale  image  in
Fig.\ref{doppler} and Fig.\ref{doppler1}.  Also in  Fig.\ref{doppler} are  displayed   trailed spectra  of
\ion{H$\beta$}{} and \ion{He}{ii} in  phase space and their corresponding
reconstructed counterparts.  Two features in the maps are distinct: an
accretion  disk seen  as a  dark circle  extending to  up  to $\sim
-700$\,km/sec on H$\beta$ doppler tomogram and a bright spots detected
in  all three  maps  to  the left  and  below the  center  of mass  at
velocities    $V_x\sim -225~km~s^{-1}$,    $V_y\sim -100~km~   s^{-1}$    in
\ion{He}{ii} 4686.  Apart  from the spot a cometary  tail linked to it
and  extending  to $V_x\approx V_y\approx  -500~km~s^{-1}$  can  be
clearly seen  on the \ion{He}{ii} map. These  we identify with  the mass
transfer  stream and  its shape  was  essential for  our selection  of
the ballistic trajectory.
A  helpful  assistance in  interpreting  Doppler  maps are  additional
inserted plots which  mark the position of the  secondary star and the
ballistic trajectory of the gas stream.  Here we used our estimates of
P$_{orb}$  and M$_2$  with various  combination  of $i$  and $q$  from
Fig.\ref{angle} in  order to  obtain the ``best  fit'' (by  simple eye
inspection)  of  calculated  stream  trajectory with  the  gray  scale
image.  Our preferred  solution of inclination is $i=35^o\deg$.
 It is   marked on  Fig.\ref{angle}  and is given  in  Table.\ref{Param}.
Of  course other  close  solutions  are applicable.
Comparing the  location of spots  in Doppler maps of  \ion{He}{ii} and
\ion{H$\beta$}{} we can distinguish actually two hot spots on the disk
(Fig.\ref{doppler}).  The  elongated spot coinciding  in both emission
lines is caused probably by the  mass transfer stream and the shock of
impact with  the disk,  while the compact  dense spot  toward negative
V$_y$'s, much better seen in \ion{H$\beta$}{} than in the  two other lines,
is   a  result  of   heating  of   the  disk   by  X-ray   beam.   The
\ion{C}{iii}/\ion{N}{iii} pattern mostly  repeats that of \ion{He}{ii}
with less intensity of course (Fig.\ref{doppler1}).

\section {Conclusions}

The \object{\wga} results to  be a classical ``textbook'' Intermediate
Polar.  It has an orbital  period P$_{orb}=4\fh36$ above the period gap, as
the  vast majority  of IPs.  It  exhibits X-ray  and optical  coherent
pulsations   of  the   order   of  $\sim0.05$P$_{orb}$,   undoubtfully
originating from  asynchronous spin of  magnetic WD in a  close binary
system.  Also the beat period  in optical light is detectable. This is
another characteristic of Intermediate Polars.

Other orbital parameters derived from assumption that the system obeys the
P$_{orb}$$\sim$M$_2$   relation  for   CVs  are   also   in-line  with
accumulated data on other IPs and theoretical aspects 
(Warner, \cite{Warner}, Paterson, 
\cite{Patterson2}, see also URL\footnote{http://lheawww.gsfc.nasa.gov/users/mukai/iphome/members.html\#cand}
).
  
The radial velocity curves, the light curve and the Doppler tomography
confirm the presence of an accretion  ring around the WD and the existence of hot
spots caused  by heating of parts of  the disk by the X-ray  beam and from
interaction with the  mass transfer stream.

\begin{acknowledgements}  
   
This work was supported in part by CONACYT project 25454-E and DGAPA
project. 
      
\end{acknowledgements}

\end{document}